
\magnification=1200
\pretolerance=10000
\baselineskip=24 pt
\noindent
\centerline {\bf Curvature energy effects on strange quark matter nucleation}

\centerline {\bf at finite density}
\vskip 1 true cm
\noindent
\centerline {J.E.Horvath}

\noindent
\centerline {\it Instituto Astron\^omico e Geof\'\i sico,
Universidade de S\~ao Paulo}

\centerline {\it Av. M.St\'efano 4200 -
Agua Funda (04301-904) S\~ao Paulo SP - Brasil}

\noindent
\centerline {and}
\noindent

\centerline {\it Department of Space Physics and Astronomy, Rice University}

\centerline {\it P.O.Box 1892, Houston, Texas 77251, U.S.A.}

\vskip 2 true cm
\noindent
{\bf Abstract}

We consider the effects of the curvature energy term on thermal strange
quark matter nucleation in dense neutron matter. Lower bounds on the
temperature at which this process can take place are given and compared
to those without the curvature term.

\bigskip
\noindent
{PACS numbers: 97.60.Jd, 12.38.Mh}

\vfill\eject
Strange quark matter (hereafter SQM) [1,2] remains an interesting and
intriguing idea which could be important for the proper
understanding of several astrophysical and cosmological processes [3].
Among them, its appearance in dense astrophysical objects has been considered
[4,5] as a pre-condition for the "burning" of the host [6,7], a phenomenon
that is still being studied and may reveal rich novel physics [8,9,10,11].
So far, the critical issue of the formation of SQM has been only
preliminary addressed to understand when and how it may happen in actual
situations [12]. We shall consider in this work the effects of the so-called
curvature term on the nucleation rate and clarify our earlier work [5] on
the subject.

One of the many possibilities of SQM appearance in dense objects [13] we
have previously addressed is the (simple) hypothesis of a thermal nucleation
[5]. Generally speaking, the knowledge of the free energy as a function of
the thermodynamic quantities is the starting point of the thermodynamic
approach and it is usual for most situations in which a nucleation takes
place just to keep the first two terms (volume and surface) of the expansion
around the bulk limit. However, for this specific problem, Mardor and
Svetitsky [14] have recently shown that the third (linear in the radius $r$)
term should also be kept in order to achieve a proper description of the
process. The appropiate form of this contribution, referred as the
"curvature" energy may be written as [14,15]

$$ E_{c} \, = \, {{g \, r} \over {3 \pi}} \int_{0}^{\infty} dk \; k \,
[ 1 + exp ((k - \mu)/T)]^{-1}  = $$

$$ = {{g \mu^{2}} r \over {6 \pi}} [ 1 + O(T/\mu)^{2}]  \eqno(1) $$

where $g$ is the statistical weight, $r$ is the radius (we assume a spherical
region) and $\mu$ is the chemical potential. Note that the first line
expression is valid for massless quarks in a M.I.T. bag [16] and the second
line is the integrated form valid for the case $T \ll \mu$, which is always
satisfied in our strongly degenerate environment.

Because of this term the problem of SQM nucleation in a proto-neutron star
[4,5] must be reconsidered to address how the previous results are affected.
Consider the work done to form a bubble of SQM (labeled as "$i$") in a
nuclear matter medium (labeled "$e$"), the system ($i + e$) assumed to be in
thermal equilibrium (common temperature $T$) but not in chemical or
mechanical equilibrium

$$ W = - {4 \over {3}} \pi r^{3} (P_{i} - P_{e}) + 4 \pi \sigma r^{2} -
8 \pi \gamma r + {4 \over {3}} \pi r^{3} n_{i} (\mu_{i} - \mu_{e}) \eqno(2) $$

where $P$ is the pressure, $\sigma$ the surface tension coefficient, $\gamma$
the curvature coeficcient and $n_{i}$ the particle number density [8].
Extremization of $W$ as a function of the radius and the knowledge of the
coefficients permits the evaluation of the critical size of the bubbles
beyond which they are able to grow at the expense of the metastable nuclear
phase; but before discussing the existence of the critical points it is
important to note that, as the coefficients $\sigma$ and $\gamma$ are
linear combinations of the contributions from each side, they should be
properly defined. Clearly $\sigma = \sigma_{i} + \sigma_{e} > 0$ is required
to avoid an unstable QCD vacuum [17], but $\gamma = \gamma_{e} - \gamma_{i}$ is
found to be negative for the problem since the expected $\gamma_{e}$ is
small [18] compared to the dominant $\gamma_{i}$. This is consistent with
earlier work [19] on the "inverse" problem of hadron bubble nucleation in
cosmological SQM nuggets. The value of $\gamma$ can be found directly by
comparing eqs.(1) and (2) and may be written as

$$ \gamma = {3 \over {8 \pi^{2}}} \mu^{2} \, \sim \, 18 \; MeV \, fm^{-1}
{\bigl( {\mu \over {300 \, MeV}} \bigr)}^{2}.  \eqno (3) $$

We shall assume hereafter that $\sigma$ and $\gamma$ are due to the SQM
contribution since the respective nuclear matter values may be included in
the uncertainties of the former due to their smallness.

Defining $\Delta P = P_{i} - P_{e}$ and $\Delta \mu = \mu_{i} - \mu_{e}$
and putting $F = -n_{i} \Delta \mu + \Delta P  > 0$ the work of eq.(2) can
be minimized with respect to the radius ; i.e. $\partial W / \partial r = 0$
to find the critical points

$$  r_{c} = {\sigma \over {F}} ( 1 \, \pm \, \sqrt {1 + \beta})
\eqno (4)  $$

where $\beta \equiv 2 F |\gamma| / \sigma^{2}$ is a positive definite quantity
that includes all the effects of the curvature term. We see that unless the
nuclear matter curvature term exceeds $\gamma_{i}$ (which is certainly not
expected) both roots are real and correspond to a maximum $r_{+}$ and a
minimum $r_{-} < 0$ of $W$. For the former

$$ W_{c} = W (r_{+}) = {4 \pi \sigma^{3} \over {3 F^{2}}}
[2 + 2 (1 + \beta)^{3/2} + 3 \beta]  \eqno (5) $$

It is the magnitude of $\beta$ which determines the increase of
the critical radius as compared to the $\gamma = 0$ case and suppresses the
nucleation rate. Typically

$$ \beta \simeq \, 6 \,
{\bigl( {F \over {20 \, MeV \, fm^{-3}}} \bigr)}
{\bigl( {\gamma \over {18 \, MeV \, fm^{-1}}} \bigr)}
{\bigl( {\sigma \over {(75 \, MeV)^{3}}} \bigr)}^{-2}   \eqno(6) $$

where we have scaled to a large value of $\sigma$ [19] attempting to address
the minimum effect to be expected from the curvature term (see also [20] for
a different approach to the calculation of $\sigma$). Even though
the actual $\sigma$ is not accurately known, our results can be easily
scaled for any value of this parameter. What is clear from eqs.(5) and (6) is
that the curvature term is indeed very important for the nucleation of SQM
in dense environments.
The idea now is to set a lower bound
on the temperature (i.e. the treshold for SQM nucleation) to see when and if
that process is possible.

In the thermodynamic approach of Refs. [8,14,21] the nucleation rate is
essentially dominated by a Boltzmann factor $exp(-W_{c}/T)$, therefore in
principle the knowledge of the thermodynamic quantities in eq.(6) (all
functions of $\mu_{i}$) for a given state of the system is possible. Since the
information contained in the growth of SQM bubbles is not thus included we
have alternatively [5] estimated the lower bound on $T$ we are interested in by
using the Fokker-Planck-Zel'dovich approach based on the kinetic equation [22]

$$ {\partial f \over {\partial t}} = \, - {\partial \xi \over {\partial t}}
\eqno(7) $$

where $f \, (r,t)$ is the time-dependent size distribution of SQM bubbles
(assumed to have a large number of particles, see below) and $\xi$ is the
nucleation rate (i.e. the number of SQM bubbles above the critical size per
unit time and volume). In that theory, the integration of eq.(7) must be
performed imposing appropiate boundary conditions and involves the evaluation
of the equilibrium size distribution function

$$ f_{o}(r_{c}) = r_{c}^{2} \, n_{i} \, n_{e} \,
exp (- 4 \pi \sigma r_{c}^{2}/ 3 T)  \eqno(8) $$

in which we have followed Ref.[22] to give an estimate of the prefactor. The
important point here is that, since we are interested in the SQM bubbles in
the neighbourhood of the critical radius $r_{c}$ (and hence corresponding to
a narrow range around the maximum $W_{c}$), a linear term such as the
curvature one does not contribute and $\xi$ can be calculated in a Gaussian
approximation, namely

$$ {1 \over {\xi}} = \, \int_{0}^{\infty}
{dr \, e^{( -4 \pi \sigma (r - r_{c})^{2} / T)} \over {B(r) f_{o}(r_{c})}}
\eqno(9) $$

where $B(r)$ is the size diffusion coefficient. After integrating eq.(9) and
imposing a physically-motivated form for $B(r_{c}) = T/ 8 \pi \tau_{w}$, with
$\tau_{w} \sim \, 10^{-9} \, s$ being the weak-interaction
time-scale in the medium (see Refs.[3,5,6,11]), we may write $\xi$ as

$$ \xi = {r_{c}^{2} \, n_{i} \, n_{e} \over {4 \, \pi \, \tau_{w}}} \,
{\bigl( {T \over {\sigma}} \bigr)}^{1/2} \, exp (- 4 \pi \sigma r_{c}^{2}/3 T)
\eqno(10) $$

in which all the effects of the curvature enter only through the value of
$r_{c}$ given in eq.(4). In order to determine the total number of bubbles that
nucleate we shall, as in Ref.[5], multiply the rate $\xi$ by the time-scale
in which favorable conditions for nucleation exists $\Delta t$
and by the volume available for the process $V_{o}$. The former is determined
by the initial cooling properties of the proto-neutron star and is generally
accepted to be $\sim$ seconds [23], therefore we shall use the fiducial
value $\Delta t = 1 \, s$. The latter is more difficult to address because it
refers to the region above an uncertain density $\rho_{o}$ in which the matter
is sufficiently compressed to undergo the transition to SQM, which is
intrinsically time-dependent since the star compacts while its lepton content
decreases due to neutrino emission [23]. We use $V_{o} = (10^{5} \, cm)^{3}$
corresponding to the inner sphere with $R < 0.1 \, R_{n}$ (where $R_{n}$ is the
radius of the object) for any instant of its evolution. Fortunately, $\Delta t$
and $V_{o}$ are not critical for our argument since the nucleation rate is
largely dominated by the exponential factor in eq.(10).

Since the condition for a neutron $\rightarrow$ strange conversion
[2,6,7,9-12] is that
at least one SQM bubble is nucleated to start it, we impose
$\xi \, \Delta t \, V_{o} \geq \, 1$, or

$$ {r_{c}^{2} \, n_{i} \, n_{e} \over {4 \, \pi}}
{\bigl( {T \over {\sigma}} \bigr)}^{1/2} \;
{\bigl( {\Delta t \over {\tau_{w}}} \bigr)} \, V_{o} \,
exp \, {\bigl( - \, 4 \, \pi \, \sigma \, r_{c}^{2}/ \, 3 \, T \bigr)} \,
\geq \, 1 \eqno(11) $$

Which shows explicitely the dependence on the various factors, particularly
the appearance of the quotient of the two relevant time-scales of the
problem and the implicit dependence on $\beta$ through $r_{c}$ given by eq.(4).
After defining $x = T/\sigma$ and replacing the numerical values eq.(11) can
be solved for $x$ to yield the treshold temperatures (those satisfying the
imposed condition) $T_{\gamma}$ shown in Table 1. We also show for comparision
the treshold temperatures corresponding to the same set of parameters and
$\gamma = 0$, denoted as $T_{o}$. Thus, the effects of the curvature term
on the nucleation rate can be quantitatively appreciated in this kinetic
approach.
As expected, the curvature term makes the minimum temperature for nucleation
to be higher, i.e. suppresses the nucleation for a given temperature.

The numbers in Table 1 may be compared to those from our former work [5],
where instead of imposing $r_{c}$ given as the critical radius of $W$ we
have tried to express it in terms of the minimum (critical) numbers of quarks
in the bubble $N_{qc}$. Although it was clear that the use of a bag-like
relationship overestimated the critical size $r_{c}$ (the external pressure
and chemical difference term $F$ did not appear in such an estimate, so that
we had assumed essentially a free configuration while the actual one is under
a large compression), it is necessary to state that the estimate was too rough
and the derived $r_{c}$ resulted too large by a factor of $\sim \, 3$.
Therefore, and due to the sensitiveness of the treshold temperatures on that
parameter, they happen to be too high by an order of magnitude and should be
corrected (see Table 1). However, the conclusions of Ref.[5] are essentially
the same as those from the present work : even though the curvature term
acts against the SQM nucleation; the physical temperature of a just-born
proto-neutron star [23] is in any model more than enough to drive an efficient
"boiling" of the neutron material. Moreover, observations of the neutrino flux
from SN1987A are consistent with an effective temperature
$T_{eff} \, \sim \, 4 \, MeV$, which may be taken as an experimental lower
limit to the actual core temperature.
Therefore, and even if  several improvements on the present
estimate may and should be attempted, the conclusion seems to be robust in
the SQM hypothesis framework and points to a prompt conversion of all
neutron stars into strange stars.

\noindent
{\bf Acknowledgements }

This work has been performed during an author's visit to the Space Physics
\& Astronomy Department, Rice University, whose hospitality is gratefully
acknowledged. The Brazilian Foundation FAPESP is also acknowledged for a
Fellowship that made possible the referred visit.

\vfill\eject

\noindent
{\bf References}

\vskip 1 true cm
\noindent
1) A.Bodmer, {\it Phys. Rev. D}{\bf 4}, 1601 (1971).

\noindent
2) E.Witten, {\it Phys. Rev. D}{\bf 30}, 272 (1984).

\noindent
3) See the {\it Proceedings of Strange Quark Matter in Physics and
Astrophysics}
, Aarhus, Denmark,1991, edited by J.Madsen and P.Haensel [{\it Nuc.Phys.B Proc.
Supp.}{\bf 24B} (1991)] for an extensive list of references.

\noindent
4) W.Slomi\'nski, {\it Acta Phys.Pol.B}{\bf 21}, 245 (1990)>

\noindent
5) J.E.Horvath, O.G.Benvenuto and H.Vucetich, {\it Phys. Rev. D}{\bf 45}, 3865
(1992).

\noindent
6) A.V.Olinto, {\it Phys. Lett. B}{\bf 192}, 71 (1987).

\noindent
7) C.Alcock, E.Farhi and A.V.Olinto, {\it Astrophys. J.}{\bf 310}, 261 (1986).

\noindent
8) M.L.Olesen and J.Madsen, {\it Phys. Rev. D}{\bf 47}, 2313 (1993).

\noindent
9) O.G.Benvenuto and J.E.Horvath, {\it Phys. Rev. Lett.}{\bf 63}, 716 (1989).

\noindent
10) J.E.Horvath and O.G.Benvenuto, {\it Phys. Lett. B}{\bf 213}, 516 (1988).

\noindent
11) H.Heiselberg, G.Baym and C.J.Pethick , in Ref. 3.

\noindent
12) J.E.Horvath, talk given at the
{\it VI J.A.Swieca School on Nuclear Physics}, Campos do Jord\~ao, Brazil,
(1993).

\noindent
13) A.V.Olinto, in Ref.3.

\noindent
14) I.Mardor and B.Svetitsky, {\it Phys. Rev. D}{\bf 44}, 878 (1991).

\noindent
15) J.Madsen, {\it Phys. Rev. Lett.}{\bf 70}, 391 (1993).

\noindent
16) The contribution of massive species to the curvature energy is not known
as yet and has been neglected here.

\noindent
17) E.Farhi and R.L.Jaffe, {\it Phys. Rev. D}{\bf 30}, 2379 (1984).

\noindent
18) F.D.Mackie, {\it Nuc. Phys. A}{\bf 245}, 61 (1975).

\noindent
19) M.S.Berger and R.L.Jaffe, {\it Phys. Rev. D}{\bf 35}, 213 (1987) ;
see also R.L.Jaffe in Ref.3.

\noindent
20) B.C.Parija, {\it Phys. Rev. C}{\bf 48}, 2483 (1993).

\noindent
21) C.Alcock and A.V.Olinto, {\it Phys. Rev. D}{\bf 39}, 1233 (1989).

\noindent
22) E.M.Lifshitz and I.Pitaevskii, {\it Physical Kinetics}
(Pergamon, London, 1980).

\noindent
23) A.Burrows, {\it Annu. Rev. Nuc. Part. Sci.}{\bf 40}, 181 (1991) ;
D.N.Schramm and J.Truran, {\it Phys. Rep.}{\bf 189}, 91 (1991).

\vfill\eject

{\bf Tables}
\vskip 2 true cm

{\bf Table 1.} Treshold temperatures for SQM nucleation (see text for details).

$$ \vbox {\settabs 5\columns

\+ $\sigma / MeV^{3}$ &  $r_{c} / fm$   & $T_{\gamma} / MeV$  &
$ r_{co} / fm$  & $T_{o} / MeV$ \cr
\+ $(65)^{3}$   &   1.74  &  0.63  &  0.71  &  0.10 \cr
\+ $(75)^{3}$   &   1.98  &  1.26  &  1.09  &  0.38 \cr} $$

\bye